\documentclass[traditabstract]{aa}
\usepackage{graphicx}
\usepackage{txfonts}
\usepackage{natbib}

\begin{document}

\title{Transmission spectral properties of clouds for hot Jupiter exoplanets}

   \author{H. R. Wakeford \inst{1}
                 \and
                 D. K. Sing  \inst{1}
                 }
   \institute{Astrophysics Group, University of Exeter, Stocker Rd, Exeter, EX4 4QL \\
              \email{hannah@astro.ex.ac.uk}
              \newline
              \newline
            Submitted 2014 May 14; Accepted 2014 Sep 25.
             }

\date{}

%
%
\abstract
{Clouds play an important role in the atmospheres of planetary bodies. It is expected that, like all the planetary bodies in our solar system, exoplanet atmospheres will also have substantial cloud coverage, and evidence is mounting for clouds in a number of hot Jupiters. To better characterise planetary atmospheres, we need to consider the effects these clouds will have on the observed broadband transmission spectra. 
Here we examine the expected cloud condensate species for hot Jupiter exoplanets and the effects of various grain sizes and distributions on the resulting transmission spectra from the optical to infrared, which can be used as a broad framework when interpreting exoplanet spectra. 
We note that significant infrared absorption features appear in the computed transmission spectrum, the result of vibrational modes between the key species in each condensate, which can potentially be very constraining. While it may be hard to differentiate between individual condensates in the broad transmission spectra, it may be possible to discern different vibrational bonds, which can distinguish between cloud formation scenarios, such as condensate clouds or photochemically generated species.
Vibrational mode features are shown to be prominent when the clouds are composed of small sub-micron sized particles and can be associated with an accompanying optical scattering slope.
These infrared features have potential implications for future exoplanetary atmosphere studies conducted with JWST, where such vibrational modes distinguishing condensate species can be probed at longer wavelengths.
}


\maketitle{}

%
%
\section{Introduction}
Where there is an atmosphere, there are clouds, or so the evidence suggests. Every planet in our solar system with a persistent atmosphere has clouds, though they are notoriously hard to define. Here we take the definition of liquid or solid aerosol particles suspended in a planet's atmosphere. Clouds are a vital part of the energy balance of a planetary atmosphere, and they can potentially play a major role in the observational structure, blocking the atmosphere beneath them and weakening any emergent spectral lines. Jupiter's atmosphere is a prime example of how clouds can shape the atmosphere of a planet, forming coloured belts and zones of dark and light bands through vertical mixing of different species, which can represent differences in cloud depth of over one planetary scale height (\citealt{evans1972}). The clouds in the atmospheres of solar system planets are, however, hugely different from what we expect to form in the atmospheres of hot Jupiters 
(e.g. \citealt{sudarsky2003}; \citealt{marley2007}; \citealt{lodders2010}). 

Hot Jupiters occupy a vastly different region of parameter space compared to the planets in our solar system, occupying higher temperatures, spanning wider temperature regimes, pressure structures, and chemical compositions. To compute the expected transmission spectrum for different cloud condensates requires us to understand a number of different processes in a planetary atmosphere. Studies have been conducted on the impact of clouds in exoplanet atmospheres (e.g. \citealt{ackerman2001}; \citealt{fortney2005}; \citealt{helling2008cloud}; \citealt{howe2012}; \citealt{marley2013}; \citealt{morley2013}) and the condensates that are expected in a wide range of temperatures from brown dwarfs to solar system bodies (e.g. \citealt{burrows1999}; \citealt{khare2001}; \citealt{lodders2003}; \citealt{cruikshank2005};  \citealt{seager2010}; \citealt{morley2012}). 

Clouds and hazes in exoplanetary atmospheres can have a strong effect on the emerging spectra. As a principle source of irradiative scattering, their presence increases the reflected flux in the visible and near-infrared regions of the spectrum (\citealt{sudarsky2003}). In addition, significant absorption features can be present (\citealt{morley2014a}).
In principle, clouds and hazes can be the result of condensation chemistry or be photochemically produced. 
The solar system giant planets are likely dominated by photochemical stratospheric hydrocarbon hazes (\citealt{Nixon2010}) produced in a similar way to tholins in the atmosphere of Titan (\citealt{Khare1984}). 
The strong UV flux on the upper atmosphere of hot Jupiter exoplanets may generally enhance photochemically generated hydrocarbon species. 
However, studies by \citet{liang2004} have found that the abundance of hydrocarbons in close-in giant planet atmospheres is significantly less than found in Jupiter and Saturn, where the high abundance of hydrocarbon aerosols results in strong absorption features shortwards of 600\,nm. The presence of non-equilibrium photochemical species with absorption in the blue gives their atmosphere a characteristic red colour (\citealt{zahnle2009}). 
The planetary albedo, for instance the observed blue albedo measurement of HD\,189733b (\citealt{evans2013}), could help differentiate between strong Rayleigh scattering dust and red tholin-like species. 

\begin{table*}
\centering  
\caption{Table of references for n and k index for a number of condensates expected to form clouds in the upper atmosphere of hot Jupiters.}
\label{nkindex}
\begin{tabular}{ccccc}
\hline
\hline
Condensate & Reference & $\lambda$ Range & Condensation & Molecular \\
~ & n, k index & ~ & Temperature$^{+}$ & Weight \\
~ & ~  &($\mu$m) & (K) & ~ \\
\hline
SiO$_{2}$ & \citet{palik1998} & 0.04 - 11 & 1725 & 60.08 \\
~ & \citet{andersen2006} & 7 - 28 & - & - \\
~  & \citet{DOCCD}* & 6.6 - 10000 & -  & - \\ 
Al$_{2}$O$_{3}$ & \citet{koike1995} & 0.3 - 150 & 1677$^{1}$ & 101.96 \\  
FeO & \citet{begemann1995} & 10 - 100 & 1650$^{4}$ & 71.79 \\ 
~ & \citet{andersen2006} & 15 - 40 & - & - \\
CaTiO$_{3}$ &  \citet{posch2003b} & 2 - 155 & 1582$^{1}$  & 135.94 \\ 
Fe$_{2}$O$_{3}$ &    \citet{DOCCD}* & 0.1 - 987 & 1566 & 159.68 \\
Fe$_{2}$SiO$_{4}$ & \citet{day1981} & 8.2 - 35 & 1443$^{4}$ & 203.77 \\
MgAl$_{2}$O$_{4}$ & \citet{DOCCD}* & 1.6 - 270 & 1397$^{1}$ & 142.26 \\ 
FeSiO$_{3}$ & \citet{day1981}  & 8.2 - 35 & 1366$^{4}$ & 131.92 \\
Mg$_{2}$SiO$_{4}$ (Fe--rich) & \citet{henning2005} & 0.2 - 445 & 1354$^{1}$ & 140.63 \\ 
Mg$_{2}$SiO$_{4}$ (Fe--poor) & \citet{zeidler2011} & 0.19 - 800 & 1354$^{1}$ & 140.63 \\ 
MgSiO$_{3}$ & \citet{egan1975a} & 0.1 - 0.4 & 1316$^{1}$ & 100.33 \\
~ & \citet{dorschner1995} & 0.5 - 80 & - & -  \\
Na$_{2}$S & \citet{morley2012} & 0.03 - 73 & 1176 & 78.04 \\
MnS & \citet{huffman1967} & 0.1 - 3 & 1139$^{2}$ & 87.00 \\
TiO$_{2}$ & \citet{kangarloo2010b} & 0.3 - 1.2 & 1125$^{2}$ & 79.86 \\
~ & \citet{kangarloo2010a}  & 1.3 - 30 & - & - \\
NaCl & \citet{palik1998} & 0.04 - 1000 & 825$^{3}$ & 58.44 \\
KCl & \citet{palik1998} & 0.02 - 200 & 740$^{3}$ & 74.55 \\
ZnS & \cite{Querry1987} & 0.2 - 167 & ~700$^{5}$ & 97.45 \\
CH$_{4}$ & \citet{martonchik1994} & 0.02 - 72 & $\sim$80 & 16.04 \\ 
C$_{6}$H$_{12}$ & \citet{anderson2000thesis} & 2.0 - 25 & 68 & 84.1 \\ 
Titan Tholins & Khare et al. (1984) & 0.01 - 0.2  & $\le$90 & $\sim$50.0 \\ %
~  & - & 1.1 - 1000 & - & - \\
~ & \citet{ramirez2002} & 0.2 - 1 & - & - \\
\hline 
\multicolumn{5}{l}{* http://www.astro.uni-jena.de/Laboratory/OCDB/oxsul.html; ~~$^{+}$ at 10$^{-3}$ bar}\\
\multicolumn{5}{l}{$^{1}$ \citet{lodders2003}, ~$^{2}$ \citet{Grossman1972}, ~$^{3}$\citet{burrows1999}, ~$^{4}$\citet{ebel2000}}\\
\multicolumn{5}{l}{$^{5}$ \citet{morley2012}} \\
\multicolumn{5}{l}{}\\
\end{tabular}
\end{table*}

Recent studies of hot Jupiters have revealed that many of the exoplanets observed in transmission have cloudy or hazy properties, with their spectra dominated by strong optical Rayleigh and\,/\,or Mie scattering from high-altitude aerosol particles (\citealt{Pont2008}, \citeyear{pont2013}; \citealt{sing2009b}, \citeyear{sing2011b}, \citeyear{sing2013}; \citealt{Gibson2013}). Clouds and hazes in the optical range effectively obscure any features from the deeper atmosphere, including pressure-broadened alkali Na and K lines and, in some cases, mute or completely cover expected water absorption features in the near infrared. Broadband transmission spectra of exoplanets, such as WASP-12b and HD\,189733b, show strong scattering in the optical to near-infrared region of the spectrum (\citealt{redfield2008}; \citealt{huitson2012}). 
HD\,189733b is one of the most extensively studied exoplanets to date, and in transmission the atmosphere is dominated by Rayleigh scattering over the whole visible range well into the infrared, with only narrow Na and K absorption line features detected above the Rayleigh slope (\citealt{pont2013}). 

Cloud composition and formation modelling for exoplanet and brown dwarf atmospheres show a number of different approaches to seed particle growth and transport. Models from Helling et al. (eg. \citealt{Helling2007}; \citeyear{Helling2009b}; \citeyear{Helling2009a}) primarily use the top-down approach, which follows seed particle growth as it drops through the atmosphere accumulating condensates and fully accounts for the micro-physics of the grain growth. 
\citet{ackerman2001} models consider the implications of downward transport of particles by sedimentation, balanced by upward mixing of vapour and condensates, in turn describing a mean global cloud in one-dimension. As a result, these two different approaches predict different cloud compositions for exoplanet atmospheres; only with more comparative studies and observations can differentiation between such models occur. 

The location and formation of cloud condensates is informed by the temperature-pressure (T-P) profile of the exoplanetary atmosphere. The altitude, and therefore pressure, at which a cloud deck will be observed is dependent on the condensation temperature as a function of pressure, where the condensation curve crosses the planetary T-P profile (\citealt{marley2013}; \citealt{morley2013}). At and around the millibar pressure range it can be assumed, to the first order, that the temperature is constant with altitude at the limb of the planet when the absence of significant inversions is assumed (\citealt{fortney2005}).

\begin{figure*}
\begin{center}
\includegraphics[width=16cm]{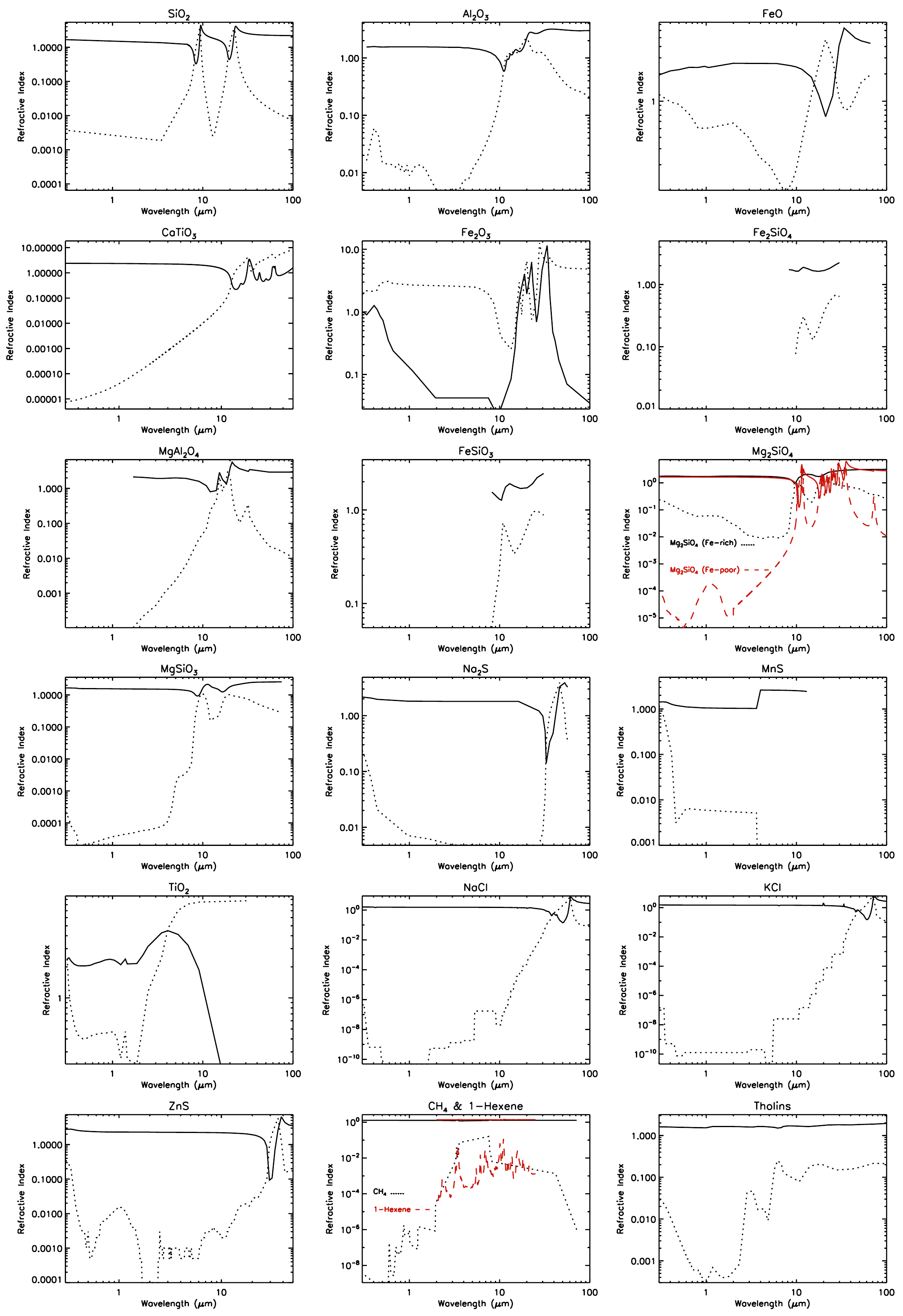}
\caption{Real (solid) and complex (dotted) index of refraction for each of the condensates listed in Table \ref{nkindex}. Condensates are listed in terms of their condensation temperature from hot in the top left corner to cold in the bottom right corner.\label{fig:nkindex}}
\end{center}
\end{figure*}

Studies of dust species in the interstellar medium (ISM) have shown that while the precise composition cannot be determined from the absorption spectra alone (e.g. \citealt{Li2001a}, \citeyear{Li2002}; \citealt{Draine2003}), it is possible to differentiate between different bond species from their stretching mode frequency, which generates strong broad absorption features at characteristic wavelengths. 
Similar detections of broadband absorption features have been made in brown dwarf atmospheric spectra (\citealt{cushing2006}, \citealt{burgasser2008}), where absorption features observed in the mid infrared are attributed to clouds of small silicate grains in the photosphere of cloudy L dwarfs.
A majority of the absorption and emission features of dust species in the ISM and brown dwarfs can be found between 3 and 25$\mu$m. These features show strong correlation to wavelengths of major optically-active vibrational modes. 

In this paper we discuss the radiative properties of multiple cloud condensates expected for hot Jupiter atmospheres and compute the expected transmission spectra over a wide wavelength regime, including well into the infrared. Condensate absorption properties rely upon the index of refraction which we discuss in \S\ref{sec:refraction} and later use to calculate the scattering and extinction cross-sections of each condensate. In \S\ref{sec:mie} we discuss Mie Theory and bond species vibrational modes in the context of potentially observable condensates. 
These are then combined to calculate the transmission spectrum in \S\ref{sec:transmission} using the planetary scale height and condensate abundance with a look at the effect of different grainsize distributions.
We apply the calculations to the well studied hot Jupiter HD\,189733b to give reference to the resulting spectra. 
In \S\ref{sec:discussion} we discuss the results in context of future James Webb Space Telescope (JWST) observations and pay particular attention to species differentiation in small grainsize condensate transmission spectra.  

%
%
\section{Radiative properties of cloud condensates} \label{sec:cloud}
Condensate chemistry in hot Jupiter atmospheres is dependent on the temperature-pressure profile and mass balance between refractory elements, such that the formation of condensate clouds at high temperatures severely depletes the gas at lower temperatures (\citealt{lodders1999}). The condensates considered in this paper are compiled from a number of sources considering equilibrium and condensate chemistry in brown dwarf and exoplanet atmospheres (e.g. \citealt{lodders1999}; \citealt{lodders2006}; Helling 2008; \citealt{morley2012}, \citeyear{morley2013}; see Table \ref{nkindex}) with consideration for cooler atmospheres like those found in our solar system (e.g. \citealt{carlson1988}; \citealt{baines1995}; \citealt{lodders2003}; \citealt{bilger2013}; see Table \ref{solarsystem}).

The condensates and cloud properties control the way that radiation moves through the planet's atmosphere, with the planetary transmission spectra dependent on the absorption and scattering of incoming and outgoing radiation. Mie theory is used to derive the absorption and scattering cross sections of solid and liquid particles (\citealt{hansen1974}), which can then be used to estimate transmission spectra. 
In this study, we do not attempt to derive a full self-consistent cloud model, but rather use simple analytic formulae and the expectations of current cloud modelling to help interpret hot Jupiter spectra. 

%
%
\subsection{Index of refraction} \label{sec:refraction}
In order to calculate all radiative properties of cloud condensates, knowledge of their refractive properties is needed. The experimental values of the refractive index need to be known so that accurate absorption cross sections can be calculated. 
Table \ref{nkindex} shows a list of cloud condensates and where the refractive indices for this work were obtained. 
The index of refraction is defined as $N=n+ik$, where $n$ and $ik$ are the real and complex parts of the refractive index, respectively. The index of refraction informs the scattering and absorption of electromagnetic waves through a material, while the complex index of refraction acts as a damping factor and is used to describe the attenuation of the waves (\citealt{liou2002}). 

Figure \ref{fig:nkindex} shows the real and complex index of refraction for each of the condensates in Table \ref{nkindex}.
The wavelength coverage of each condensate is determined by the experimental data presented in the associated papers. The calculated spectra are also dependent on the resolution of the measurements recorded. 
Features in the absorption properties of each condensate will have strong implications on the resulting spectrum of that particle. By referring to the complex index of refraction, trends emerge between different species of condensates, which can be explained by the vibrational properties of each molecule (see \S\ref{sec:modes}).

%
%
\subsection{Mie theory} \label{sec:mie}
Mie theory is an analytical solution to Maxwell's equations, which describes how to calculate the phase functions and absorption and scattering cross sections of solid or liquid particles. To compute these we use \emph{bhmie}\footnote{http://www.met.tamu.edu/class/atmo689-lc/bhmie.pro}, an IDL routine that uses Bohren-Huffman Mie scattering to calculate scattering and absorption by a homogeneous isotropic sphere (\citealt{bohren1983}). When the radius of the particle greatly exceeds the wavelength, the theory tends to geometric optics, while if the wavelength greatly exceeds the radius of the particle Mie theory tends to Rayleigh scattering as observed in a number of exoplanetary atmospheres in the UV and optical (\citealt{lecavelier2008}; \citealt{sing2011b}). Mie theory is used to derive solutions for spherical particles and is able to provide a first-order description of optical effects in non-spherical particles like those likely to exist in planetary atmospheres.

\begin{table}
\centering
\caption{Table of cloud condensates found in the atmosphere of solar system planets. }
\label{solarsystem}
\begin{tabular}{cccc}
\hline
\hline
Condensate & Planet &  Condensation  & Pressure Range$^{+}$ \\
~ & ~ & Temperature$^{+}$ & (cold-hot) \\
~ & ~~ & (K) & (bar)\\
\hline
H$_{2}$O & J, S, U, N &  274 - 348 & 4.85 - 526 \\
NH$_{3}$ & J, S, U, N &  147 - 163 & 0.66 - 7.62 \\
NH$_{4}$SH & J, S, U, N & 209-237 & 2.01 - 42 \\
H$_{2}$S & S, U, N & 116 - 124 & 0.66 - 3.23 \\
CH$_{4}$ & U, N & $\sim$80 & 0.94 - 1.2 \\ 
\hline
Titan tholins & Titan & $\le$90 & ~ \\ 
\hline
\multicolumn{3}{l}{$^{+}$\citet{carlson1988}}
\end{tabular}
\end{table}

%
%
\subsubsection{Scattering and extinction efficiency} \label{sec:efficiency}
Given the index of refraction, we can calculate the extinction, scattering, and absorption cross-section ($\sigma_{abs}$) and efficiency for a given particle size, where the scattering efficiency ($Q_{scatt}$) and extinction efficiency ($Q_{ext}$) are
    \begin{equation}
        Q_{scatt} = \frac{\sigma_{scatt}}{\pi a^{2}} = \frac{2}{x^2}\sum_{n=1}^{\infty} (2n + 1) [|a_{n}|^{2} + |b_{n}|^{2}] \rm{\,, and}
    \end{equation}

    \begin{equation}
        Q_{ext} = \frac{\sigma_{ext}}{\pi a^{2}} = \frac{2}{x^{2}}\sum_{n=1}^{\infty} (2n +1) R_{e} (a_{n} + b_{n}) \rm{\,,}
    \end{equation}

\noindent where $\sigma_{ext}$ and $\sigma_{scatt}$ are the extinction and scattering cross-sections respectively, $a$ is the grain radius, $x$ is the size parameter ($x = 2 \pi a / \lambda$), and $a_{n}$ and $b_{n}$ are the mix coefficients expressed in terms of the complex index of refraction (see \citealt{sharp2007} for full equations). From this the extinction, scattering, and absorption cross section can be calculated as
    \begin{equation}
        \sigma_{ext} =  \sigma_{scatt} + \sigma_{abs} \ _{.}
    \end{equation}

%
%
\section{Transmission spectrum} \label{sec:transmission}

\begin{figure*}
\centering
\includegraphics[width=16.7cm]{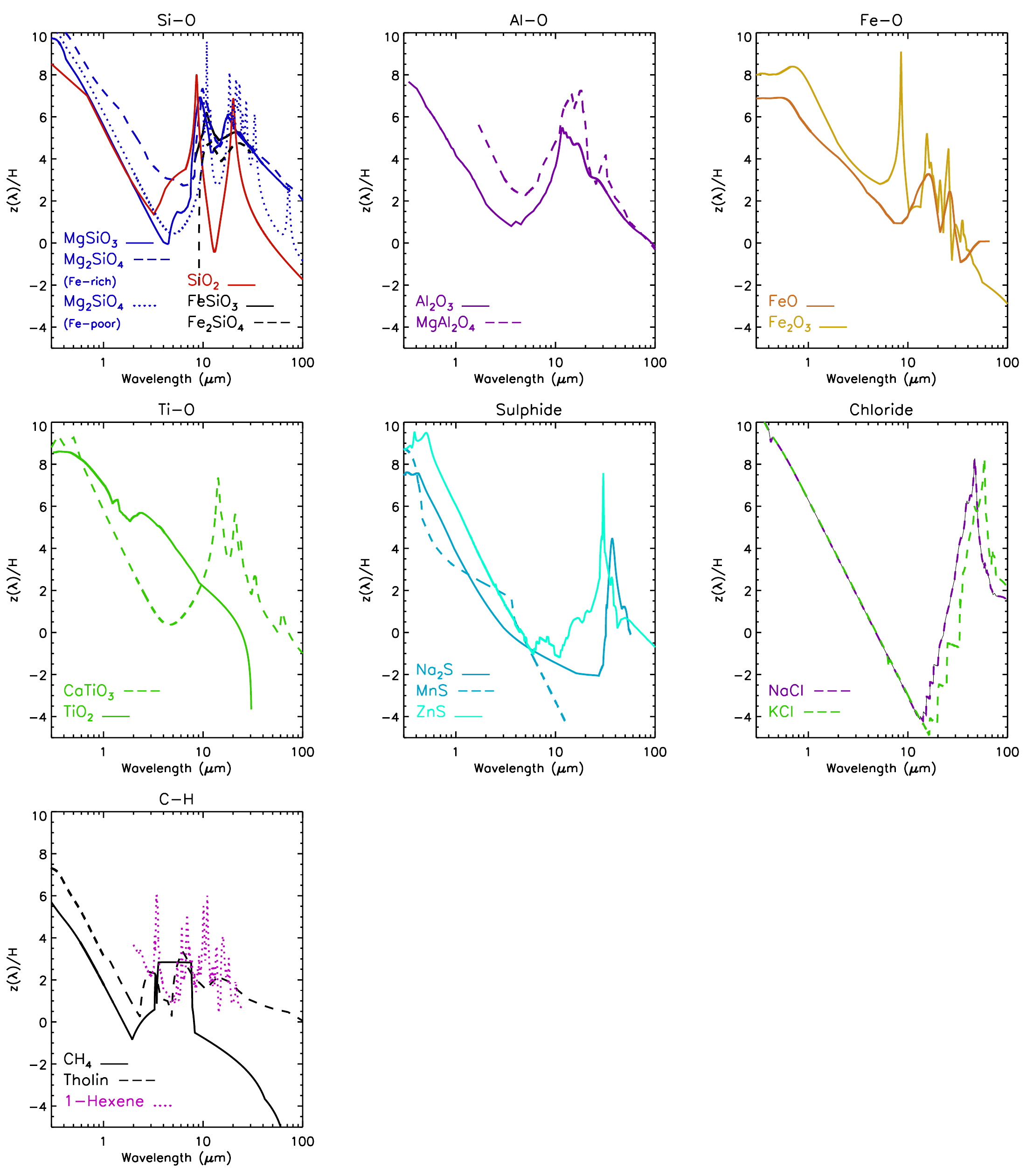}
\caption{Transmission spectra for a number of condensates expected in hot Jupiter atmospheres using HD\,189733b system parameters (a = 0.1$\mu$m, T$_{eff}$=1350\,K). The condensates have been separated out according to their primary bond where vibrational modes between these species dominate the spectra. \label{fig:bond_plots}}
\end{figure*}

To compute the transmission spectrum for a given condensate we calculate the effective altitude, $z(\lambda)$, of the atmosphere as a function of wavelength using the analytical formula of \citet{lecavelier2008} and the cross sections estimated from Mie scattering. This needs to be done separately for each of the condensates considered as it relies upon the planetary scale height and specific abundance for each cloud considered. 
Here we use the well-studied hot Jupiter HD\,189733b as an example atmosphere with R$_{p}$=\,1.138\,R$_{\rm J}$, R$_{*}$=\,0.756\,R$_{\sun}$, g$_{p}$=\,3.34\,ms$^{-2}$, and T$_{eff}$\,=\,1350\,K  (\citealt{southworth2010}; \citealt{torres2008}).
The effective altitude is given by

    \begin{equation}
      z(\lambda) = H\, ln \left( \frac{\xi_{abs} P_{(z=0)}\sigma_{ext}(\lambda)} {\tau_{eq}} \times \sqrt{\frac{2 \pi R_{p}}{k_{B}T\mu g}}\, \right) \rm{\,,}
    \end{equation}

\noindent where $H$ is the planetary scale height, $\xi_{abs}$ is the abundance of the dominant species, $P_{z=0}$ is the reference pressure at $z=0$, $\sigma_{ext}(\lambda)$ is the wavelength dependent extinction cross-section, $\tau_{eq}$ is the equivalent optical depth at the measured transit radius ($\sim$0.56; \citealt{lecavelier2008}), and $k_{B}$ is the Boltzmann constant.

A planetary scale height, $H$, is the altitude range over which the atmosphere pressure decreases by a factor of e, such that
    \begin{equation}
        H = \frac{k_{B}T}{\mu_{m}m_{H} g} \rm{\,,}
    \end{equation}

\noindent where $T$ is the estimated atmospheric temperature, $m_{H}$ is the mass of a hydrogen atom, $\mu_{m}$ is the mean molecular weight of the atmosphere, and $g$ is the surface gravity. 
For condensates their scale height, $H$, can potentially be smaller than the gaseous scale height; $H \sim H_{g}/3$ (\citealt{fortney2005}). \citet{lecavelier2008} showed that, in the case of HD\,189733b, the gaseous pressure scale height can be equated to that of the condensate scale height determined from the observed Rayleigh slope, which implies strong vertical mixing of condensates in the planetary atmosphere. We use the same assumption here.
We compute the condensate cloud transmission spectra for an isothermal atmosphere. Following temperature-pressure profiles from Fig. 1 of \citet{showman2008} and Fig. 5 of \citet{fortney2010} it can be seen that at altitudes probed in slant geometry at the terminator, in the mbar pressure range, the transmission spectra are not overly sensitive to changes in temperature, where $\Delta$T is of the order of 100\,K (also see \citealt{howe2012}).

%
\begin{table}
\centering
\caption{Table of vibrational modes for the major diatomic bond species in the different cloud condensates considered in this paper.}
\label{bond_freq}
\begin{tabular}{cccc}
\hline
\hline
Major & Reduced  & Vibrational  & Wavelength, \\
Bond & Mass, $\mu_{M}$ & Frequency, $\nu$ & $\lambda$ \\
~ & (g) & (cm$^{-1}$) & ($\mu$m) \\
\hline
Si - O & 10.192 & 1110 - 830$^{a}$ & 9 - 12  \\
Al - O & 10.043 & 1100 - 350$^{c}$ & 9 - 28.7 \\
Fe - O & 12.436 & 790$^{b}$ & 12.5 \\
Ti - O & 11.99 & 850 - 150$^{d}$ & 16 - 66 \\
MnS & 20.247 & 295-220$^{e}$ & 20.2 \\
ZnS & 21.51 & 464$^{f}$ & 21.5  \\
NaCl & 13.95 & 366$^{g}$ & 13.95  \\
KCl & 18.60 & 281$^{g}$ & 18.6  \\
C - H & 0.923 & 3032$^{a}$ & 3.3 \\
\hline
\multicolumn{4}{l}{$^{a}$\citet{glassgold2012}; ~$^{b}$\citet{lehnert2002}} \\
\multicolumn{4}{l}{$^{c}$\citet{Saniger1995109}; ~~ $^{d}$\citet{gillet1993}} \\
\multicolumn{4}{l}{$^{e}$\citet{batsanov1969}; ~$^{f}$\citet{kroger1954}} \\
\multicolumn{4}{l}{$^{g}$\citet{rice2004}} \\
\end{tabular}
\end{table}

%
%
\subsection{Condensate abundance} \label{sec:abundance}
To determine the expected abundance of a certain condensate we first assume that it relies upon the metallicity abundance of the main atom from solar abundances (\citealt{burrows1999}). Here we use the example of MgSiO$_{3}$ to demonstrate the calculation (\citealt{lecavelier2008}),
    \begin{equation}
        \xi_{MgSiO_{3}}\ =\ \frac{3\ a^3\ N_{A}\ m_{p}\  \mu_{MgSiO_{3}}\ \xi_{Mg}}{2\ \pi\ \rho_{MgSiO_{3}}}
    \end{equation}

\noindent where $\xi_{Mg}$ is the solar abundance of Mg, $\rho_{MgSiO_{3}}$ is the density of MgSiO$_{3}$, and $N_{A}m_{p}$ is Avogadro's constant multiplied by the mass of a proton acting as a scaling factor.

The effective altitude, $z(\lambda)$, is then added to the bulk planetary radius to compute the observable transmission spectrum, R$_{p}$($\lambda$)/R$_{*}$, of the condensate. 

Figure \ref{fig:bond_plots} shows the transmission spectrum for each cloud condensate in Table \ref{nkindex} computed for the atmosphere of HD\,189733b given a condensate grain size of 0.1\,$\mu$m and abundances 1x solar. We plot the spectrum in units of scale height, $H$, such that the transmission spectra of different exoplanets will appear very similar, as $z/H$ is only weakly dependent upon specific values of R$_{p}$ or $g$. The different condensates have been separated into groups of their primary diatomic bond to highlight the similarities between different condensate spectra when the absorption is dominated by one vibrational state. 

%
%
\subsection{Vibrational modes} \label{sec:modes}
Major dust spectral features are determined by vibrational modes. Silicate dust has a major feature at 10\,$\mu$m, while hydrocarbons have major features at 3\,$\mu$m. The vibrational frequency, $\nu$, for the major dipole bonds considered here can be estimated assuming a harmonic oscillation with,
      \begin{equation}
          \nu = \frac{1}{2\pi c}\sqrt{\frac{K}{\mu_{M}}}
      \end{equation}

\noindent where $c$ is the speed of light, $K$ is the force constant of the bond considered, and $\mu_{M}$ is the reduced mass in grams $\mu_{M} = (m_{1} m_{2})/(m_{1}+m_{2})$.

The calculated vibrational modes and their corresponding wavelength ranges for each of the major bond species considered in this paper can be seen in Table \ref{bond_freq}. 
Unlike gaseous molecules, where the rotational structure of the molecules can be observed as individual narrow absorption lines, solid molecules suppress the rotational structure as they cannot rotate freely resulting in a smearing of the absorption lines into broad peaks. Figure \ref{fig:bond_plots} displays the transmission spectra calculated using Mie theory, and the commonalities between different condensates with the same major vibrational modes due to the main diatomic bond can be seen. 

For simple diatomic molecules like MnS and ZnS or NaCl and KCl the slight difference in vibrational modes can be seen clearly in the sharp absorption features of their transmission spectra. 
More complex molecules show broad absorption features across a range of wavelengths centred around the vibrational wavelength of the major dipole.

%
\subsection{Particle size and distribution} \label{sec:grainsize} 
The transmission spectrum for all condensate species is highly dependent on the size ($a$) of the particles composing the cloud. To loft particles to the upper atmosphere, the expectations of vertical mixing need to be taken into account. Models from \citet{parmentier2013} and \citet{heng2013} show that strong vertical mixing can keep micron or sub-micron sized particles aloft in the atmosphere where grain sizes between 0.001 and 100\,$\mu$m were considered. 
\citet{lecavelier2008} show that in the Rayleigh regime, the cross section is proportional to $a^6$. This makes the scattering and resulting transmission spectrum largely dependent on the largest grain size in the particle distribution of the cloud.

\begin{figure}
\includegraphics[width=9cm]{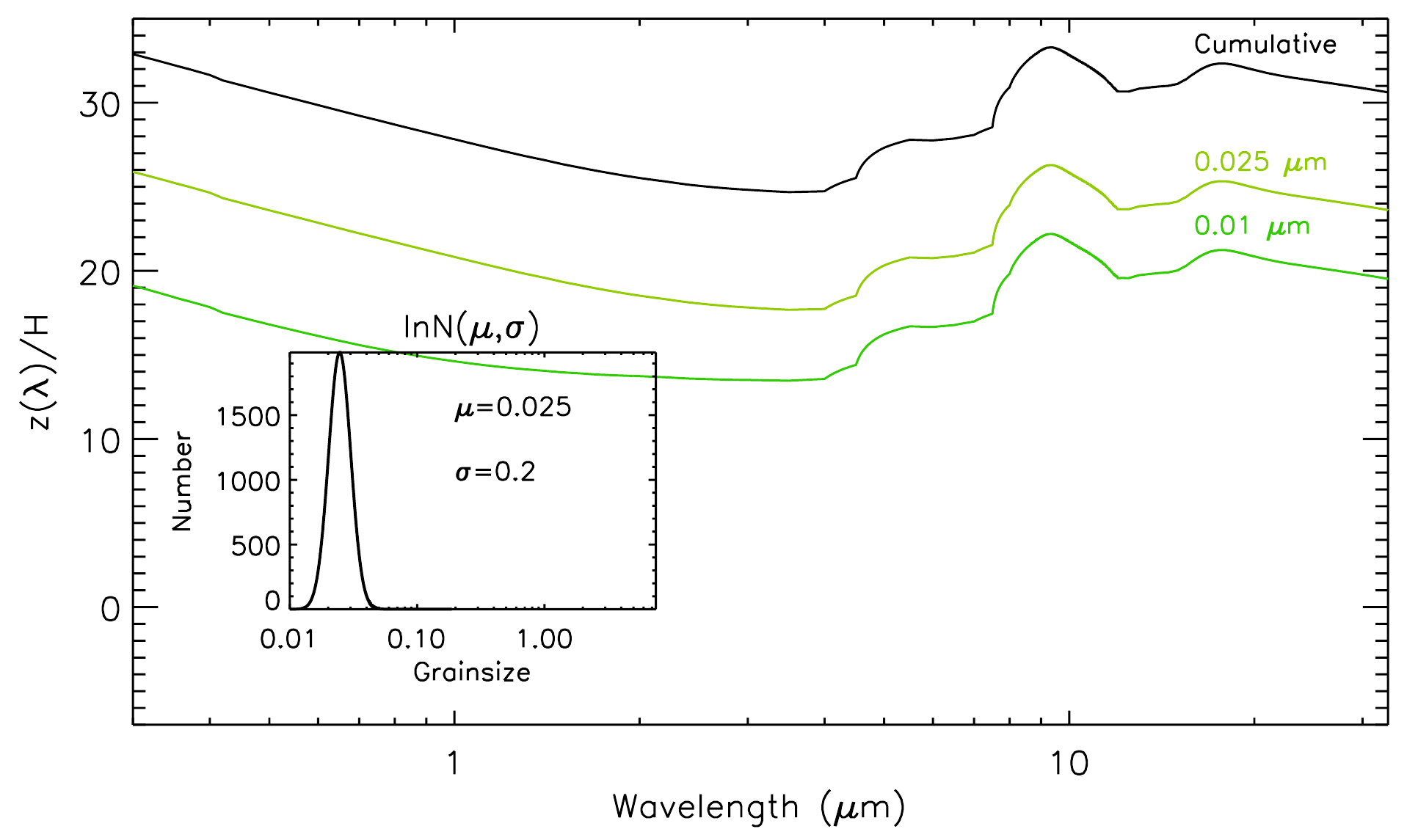}
\caption{Cumulative transmission spectrum of a log-normal distribution centered at 0.025$\mu$m with a width of 0.2 showing each of the individual spectra contributing to the final transmission spectra of the cloud. The sub-plot shows the distribution used for this cloud with the x-axis on a log scale. 
\label{fig:mu0025_sig02}}
\end{figure}

\begin{figure}
\includegraphics[width=9cm]{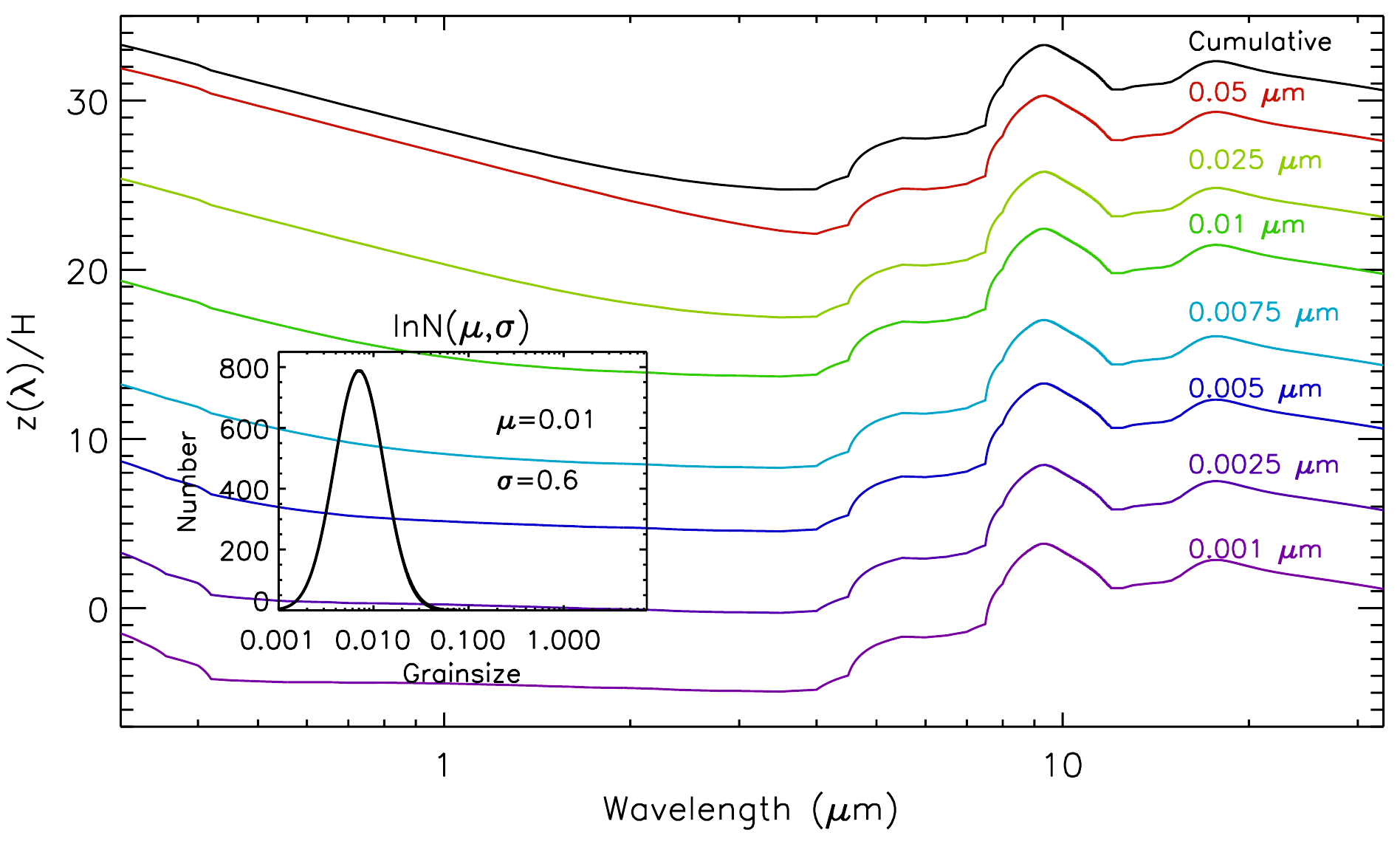}
\caption{Cumulative transmission spectrum of a log-normal distribution centered at 0.01$\mu$m with a width of 0.6 showing each of the individual spectra contributing to the final transmission spectra of the cloud. The sub-plot shows the distribution used for this cloud with the x-axis on a log scale.  
\label{fig:mu001_sig06}}
\end{figure}

\begin{figure}
\includegraphics[width=9cm]{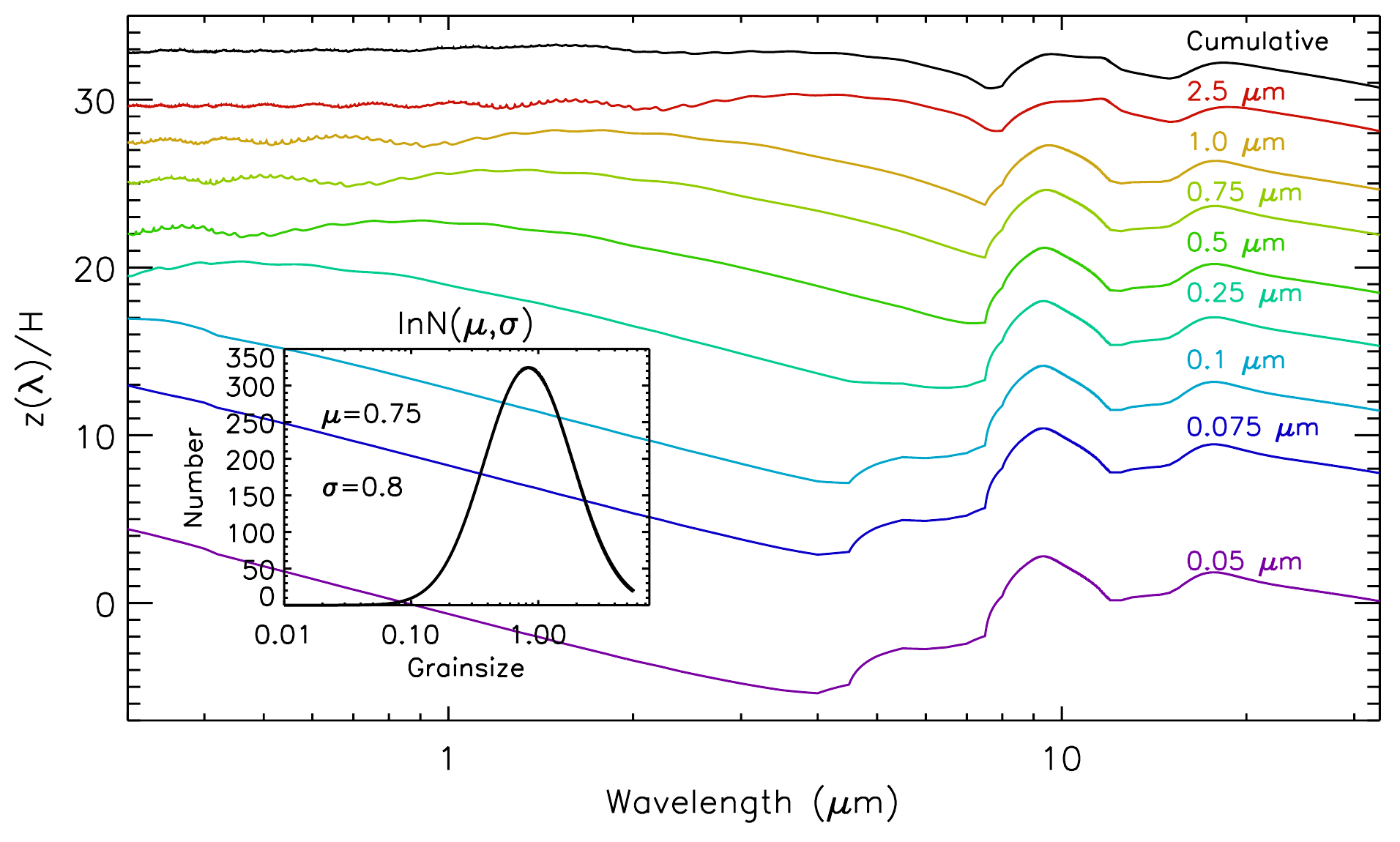}
\caption{Cumulative transmission spectrum of a log-normal distribution centered at 0.75$\mu$m with a width of 0.8 showing each of the individual spectra contributing to the final transmission spectra of the cloud. The sub-plot shows the distribution used for this cloud with the x-axis on a log scale. 
\label{fig:mu075_sig08}}
\end{figure}

To demonstrate the effect of larger grain sizes in clouds with particle distributions,
we applied a series of log-normal grain size distributions to our cloud particles and calculated the resulting transmission spectrum. Log-normal distributions ($lnN(\mu,\sigma)$) are dependent upon the centre ($\mu$) of the distribution and the width ($\sigma$). We set a grid of log-normal distributions with $\mu$\,=\,0.001\,-\,7.5\,$\mu$m and $\sigma$\,=\,0.05\,-\,1.0 and computed the cumulative transmission spectrum for the condensate clouds of each distribution. Figures \ref{fig:mu0025_sig02}, \ref{fig:mu001_sig06}, and \ref{fig:mu075_sig08} show three distributions with the contributing grain sizes for the transmission spectra, and the resulting cumulative spectrum. 
These figures demonstrate the effect of larger grain sizes on the cumulative transmission spectrum, where a small number of large grain size particles dominates the resulting cloud spectra. For distributions with only sub-micron sized particles, vibrational mode absorption features in the infrared can be seen. As evident in Figs. \ref{fig:mu0025_sig02}, \ref{fig:mu001_sig06}, and \ref{fig:mu075_sig08} all the cumulative distribution transmission spectra can be well approximated by a single grain size which is the largest in the distribution, as expected given the $\sigma\propto$\,a$^6$ relation. Distributions with sizes larger than $\sim$1\,$\mu$m tend toward completely flat, featureless spectra in the optical and infrared. Figure \ref{fig:mu075_sig08} shows the transition from prominent absorption features visible in sub-micron sized particle spectra to flat spectra where the grain sizes become larger than $\sim$1\,$\mu$m. 

Key features in hot Jupiter transmission spectra can be used as diagnostic tools to set limits on the grain size of observed clouds, and predict the likelihood of features being observed at longer wavelengths.
The presence of a Rayleigh slope in the optical part of the spectra, like that seen in HD\,189733b (\citealt{pont2013}), indicates that any clouds present in the atmosphere at transmission spectral altitudes are likely made of sub-micron sized grain particles. Further evidence of infrared features (\citealt{mccullough2014}), along with the general planetary parameters, can be used to diagnose the probability of observing condensate absorption features in the mid infrared where a majority of the condensate vibrational modes are observed.
However, if larger grain sizes are lofted up in the atmosphere to altitudes probed by transmission spectra it is unlikely significant optical to infrared features will be observed, such as in the flat transmission spectra of HAT-P-32b (\citealt{Gibson2013}).

%
%
\section{Discussion} \label{sec:discussion}

A majority of current exoplanet spectra are constructed from ground-based measurements with additional low resolution spectra from HST at wavelengths short of 1.7\,$\mu$m, and in the infrared from Spitzer (e.g. \citealt{Gibson2013}; \citealt{pont2013}; \citealt{sing2013}; \citealt{nascimbeni2013}). At present, these observations focus on the optical and near-infrared regions of the spectrum, revealing information on the portion of transmission spectra for aerosols where only scattering features are seen (e.g. \citealt{pont2013}; \citealt{sing2013}; see Fig. \ref{fig:189_h2o}). We have used Mie theory, and the expectations of current cloud modelling, to compute approximate hot Jupiter transmission spectra from the optical to far infrared regime. When interpreting observations, the slope of spectra in the optical regime is proportional to the temperature of the atmosphere and can be indicative of specific species when small grain sizes are considered. Additionally, absorption features in the near- and mid-infrared spectra can be identified as the vibrational modes of the major bond pair in the condensates considered, providing composition information.  

Models of brown dwarf atmospheres suggest that silicates are likely to form the dominant cloud structures in hot Jupiter atmospheres where temperatures are greater than $\sim$1000\,K, with sulphide clouds becoming dominant at temperatures below 900\,K at which chlorides also begin to condense out (\citealt{morley2012}).  
It is expected that hotter atmospheres could have a greater abundance of Al$_{2}$O$_{3}$ and Fe$_{2}$O$_{3}$, as silicates will not have condensed out. Additionally, Al$_{2}$O$_{3}$ and CaTiO$_{3}$ will not be present at the mbar pressure level in atmospheres with T$_{eff}\le$\,1600\,K as Al and Ca get locked up in magnesium oxides in deeper layers of the atmosphere. 

This work shows that while it is unlikely that we will be able to distinguish between individual silicate dust species, similar to the ISM and brown dwarfs, we may be able to discern a contrast between separate dust sub-classes like those shown in Figure \ref{fig:bond_plots}. As a result, an observational distinction can be placed on photochemically generated species, such as hydrocarbons with a dominant C--H bond, and condensation chemistry produced molecules like those with a dominant Si--O bond. 
Along with wavelength differentiation between major species' vibrational modes, there is also a significant altitude distinction in the transmission spectra where features in the near infrared can extend above the optical slope. At these wavelengths, predominant cloud absorption features could compete with H$_{2}$O and other molecules in the near infrared, potentially obscuring expected atmospheric features. 

%
%
\subsection{Interpreting hot Jupiter transmission spectra}

Presently exoplanet spectra are limited to the optical and near-infrared regime below 1.7$\mu$m with intermittent wavelength coverage into the infrared with Spitzer. Current exoplanet broadband transmission spectra that have evidence for clouds show commonalities in the optical regime, where all scatterers appear to be very similar.
There is growing evidence for differences in their molecular transmission spectral signatures. For instance, WFC3 spectra have detected large H$_{2}$O features in HAT-P-1b and WASP-19b (\citealt{wakeford2013}; \citealt{huitson2013}) while these features can be muted or even absent for other planets like WASP-12b and WASP-31b (\citealt{sing2013}; \citealt{sing2014}). 

The hot Jupiter HD\,189733b has been extensively studied into the infrared with observations by cold Spitzer at 3.6, 4.5, 5.8, 8.0, and 24\,$\mu$m (\citealt{knutson2007}, \citeyear{knutson2012}).  Measurements of HD\,189733b in the infrared hint at the presence of molecular absorption by water in the planet's upper atmosphere with additional evidence from both high resolution spectroscopy and eclipse spectral data (\citealt{grillmair2008}; \citealt{birkby2013}; \citealt{pont2013}; \citealt{mccullough2014}). 
We use the HD\,189733b transmission spectrum as an example hot Jupiter atmosphere, and discuss potential spectral features with regards to the condensates and photochemical species shown in this paper. Absorption from gaseous species such as H$_2$O, CO, and CH$_4$ can be present in the infrared and obscure condensate features. For HD\,189733b, optical scattering can be seen to high altitudes spanning $\sim$7\,H with infrared data giving strong constraints on the altitude levels of the gaseous molecular species such as H$_2$O. 
While only a subset of the condensates considered here are appropriate for HD\,189733b, given the planetary parameters, we use it to illustrate where other condensates could be detectable in hotter or cooler exoplanets, given their vibrational modes. 

Using the HD\,189733b transmission spectra as an example hot Jupiter we address two questions; what transmission spectral features for the various condensates can we observe and at what wavelength are we likely to observe them?

\begin{figure}
\includegraphics[width=9cm]{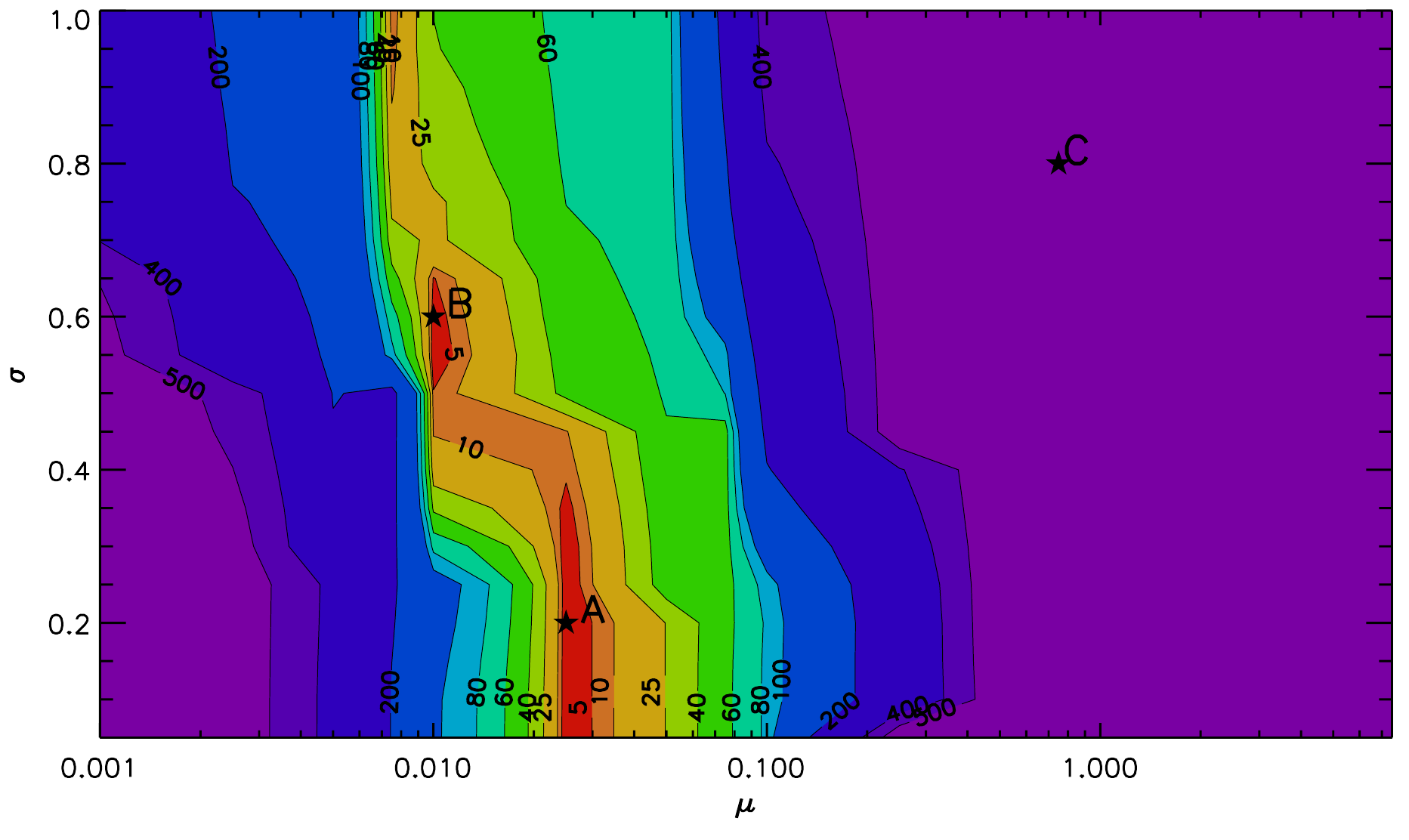}
\caption{$\Delta\chi^{2}$ grid for a series of log-normal grainsize distributions when fitted to the HD\,189733b transmission spectral data. The best fit occurs for distributions which contain $\sim$0.025\,$\mu$m size particles. The distributions in figures \ref{fig:mu0025_sig02}, \ref{fig:mu001_sig06}, and \ref{fig:mu075_sig08} are labeled A, B, and C respectively \label{fig:chi_squared_fit}}
\end{figure}

%
%
\subsubsection{Grainsize} \label{subsec:grainsize}

We fit the HD\,189733b transmission spectrum with a grid of log-normal distributions applied to the condensate MgSiO$_{3}$. We fit the cumulative spectrum of each grid point with the complete HD\,189733b spectrum by allowing only the altitude to vary. Figure \ref{fig:chi_squared_fit} shows the $\Delta\chi^{2}$ fit for each log-normal particle distribution, where the position of the expanded distributions in Fig. \ref{fig:mu0025_sig02}, \ref{fig:mu001_sig06}, and \ref{fig:mu075_sig08} are labeled A, B, and C respectively. The best fit distribution is shown to be 0.025\,$\mu$m when $\sigma$ is small, with wider distributions where the maximum grain size is 0.025$\mu$m also providing good fits to the data.

Using the best-fitting condensate model we can rule out the presence of particles larger than $\sim$0.025\,$\mu$m in the atmosphere of HD\,189733b, which would generate a `flat' spectrum from the optical through the infrared and likely hide the deeper H$_{2}$O features observed in the near infrared. 
Expanding these results to other hot Jupiters, a similar exploration of the grain size distributions which are compatible with optical and near-infrared transmission spectra can be used to predict potential condensate vibrational mode features in the infrared. These findings suggest such infrared vibrational modes will only be observable in transmission spectra where optical scattering is present with sub-micron size particle distributions. 

Given the largest grain-sized particle in the distribution produces a reasonable approximation of the transmission spectrum's shape, as shown by Figs. \ref{fig:mu0025_sig02}\,--\,\ref{fig:chi_squared_fit}, for simplicity a single grain size is assumed in subsequent sections. 

\begin{figure*}
\begin{centering}
\includegraphics[width=15cm]{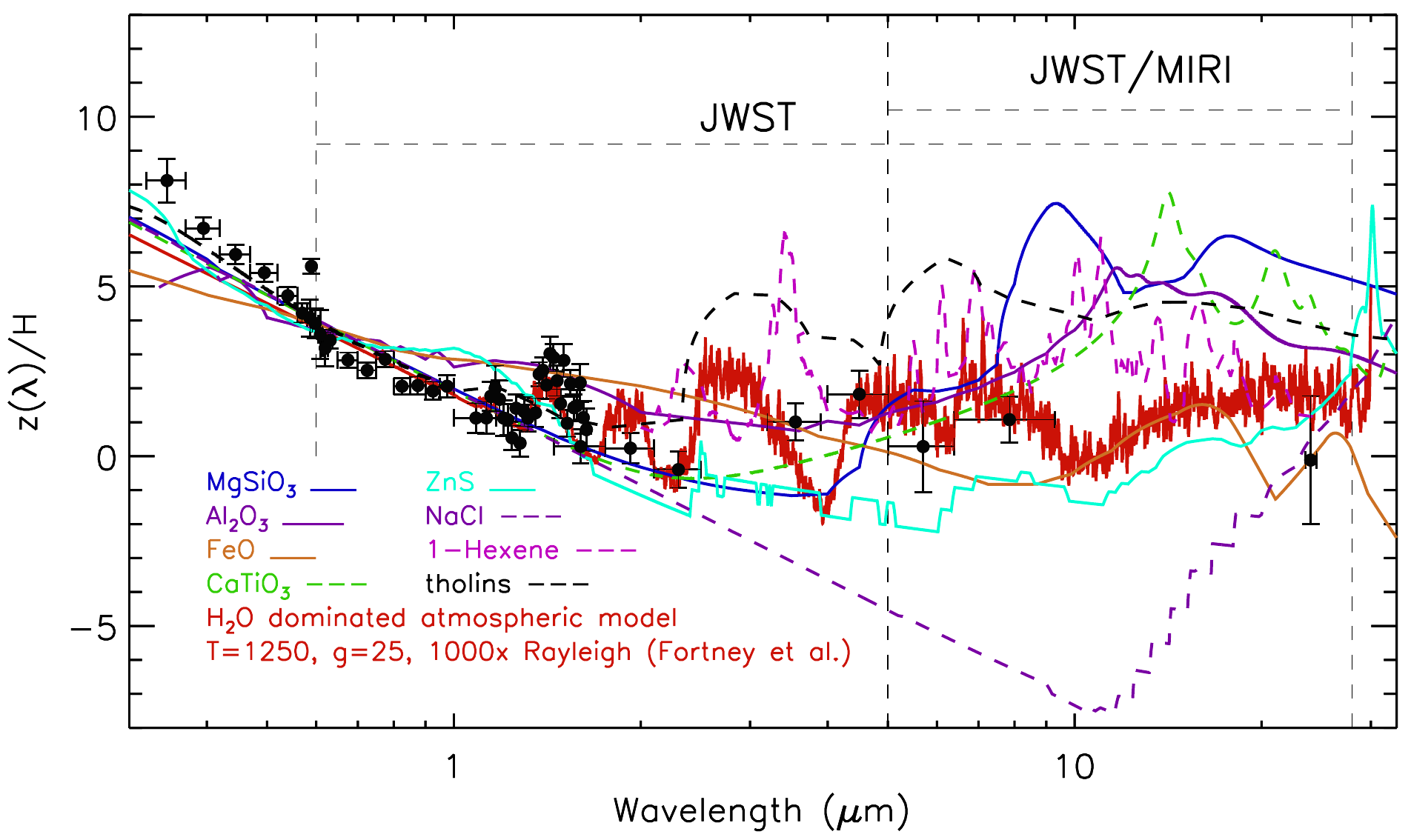}
\caption{Transmission spectrum of HD\,189733b (\citealt{pont2013}; \citealt{mccullough2014}) compared to a number of different condensates with a best fit grainsize of 0.025$\mu$m and a H$_{2}$O dominated atmospheric model from Fortney et al.}
\label{fig:189_h2o}
\end{centering}
\end{figure*}

%
\subsubsection{Condensate spectra}

We fit single grain size condensate cloud spectra from 0.001\,--\,10\,$\mu$m calculated using Mie theory for each of the condensates in Table \ref{nkindex} to the HD\,189733b data from 0.3\,--\,1.0\,$\mu$m of \citet{pont2013} where significant Rayleigh scattering is observed.

Figure \ref{fig:189_h2o} shows the observed HD\,189733b transmission spectrum from 0.3 to 24\,$\mu$m (\citealt{pont2013}; \citealt{mccullough2014}) with a representative condensate spectrum for each of the major diatomic vibrational modes shown in Fig. \ref{fig:bond_plots}.
By allowing only the altitude to vary, we find a best-fit grain size of 0.025\,$\mu$m for each of the condensates matching the grain size distribution fits shown in Fig. \ref{fig:chi_squared_fit}.

The cloud model spectra shown in Fig. \ref{fig:189_h2o} demonstrate that while scattering by aerosols becomes negligible at longer ($\sim$2\,$\mu$m) wavelengths, as the cross-section becomes small, significant infrared absorption features appear. 
In some cases, these vibrational absorption features are seen to rise to altitudes in the transmission spectra above the optical scattering levels, making them potentially detectable.
Additionally, a number of the absorption features span multiple scale heights becoming visible above the expected molecular bands of the abundant gaseous species (H$_2$O, CH$_4$). We use a \citet{fortney2010} model of HD\,189733b which has a 1000$\times$ enhanced scattering to estimate the amplitude and relative location of molecular features in the transmission spectrum, fitting for the HD\,189733b transmission data (\citealt{pont2013}; \citealt{mccullough2014}) by allowing only the model altitude to vary. The \citet{fortney2010} model is a good fit to the optical slope, the 1.4\,$\mu$m H$_2$O feature, and the Spitzer data, though only contains an artificial pure-Rayleigh scattering component, which has no effect on the transmission spectrum longer than $\sim$2\,$\mu$m.  
As evident in Figs. \ref{fig:bond_plots} and \ref{fig:189_h2o}, the scattering properties of each condensate considered here are expected to be similar shortward of 3\,$\mu$m, as absorption features, due to the vibrational modes of the molecules, are not observed in the optical and near infrared (see Table \ref{bond_freq}). 
Here we consider each of the major bond species in turn and discuss the absorption features which may be detectable for sub-micron size grains.

Si--O: Each of the silicates considered to be present in the atmospheres of hot Jupiters show a strong Rayleigh slope in the optical up to 3$\mu$m with an exception of Fe--rich Mg$_{2}$SiO$_{4}$ which shows additional absorption to that of standard Rayleigh. Major vibrational mode absorption features emerge from 9\,--\,12\,$\mu$m spanning multiple scale heights matching the altitude at the top of the optical Rayleigh slope. In Fig. \ref{fig:189_h2o} these features are represented by the transmission spectrum of MgSiO$_{3}$ which has an absorption feature across several microns reaching above the 1000$\times$ Fortney spectra. If silicate clouds dominated by small particle sizes are present in hot Jupiter atmospheres, it is likely that their presence and pressure altitude can be determined by transmission spectral features observed in the infrared by JWST/MIRI.

Al--O: Aluminium oxides also show distinct Rayleigh properties in the optical, with broadband absorption features extending from 9\,--\,28\,$\mu$m. While it is unlikely that the atmosphere of HD\,189733b contains Al--O condensate clouds, it is possible that hotter Jupiters, such as WASP-12b which has a significant optical slope, can show Al--O condensate features. As evident in Fig. \ref{fig:189_h2o}, the peak of the Al--O vibrational mode feature forms a broadband absorption spectrum between that of the two silicate absorption features at 10\,--\,12\,$\mu$m. It is possible that if Si--O or Al--O condensate clouds are present in a hot Jupiter atmosphere, such as the high temperature condensates SiO$_{2}$ or Al$_{2}$O$_{3}$, either could be identified where the altitude of the obscuring feature can exceed that of potential gaseous molecules from $\sim$8\,--\,20\,$\mu$m. 

\begin{figure*}
\begin{centering}
\includegraphics[width=15cm]{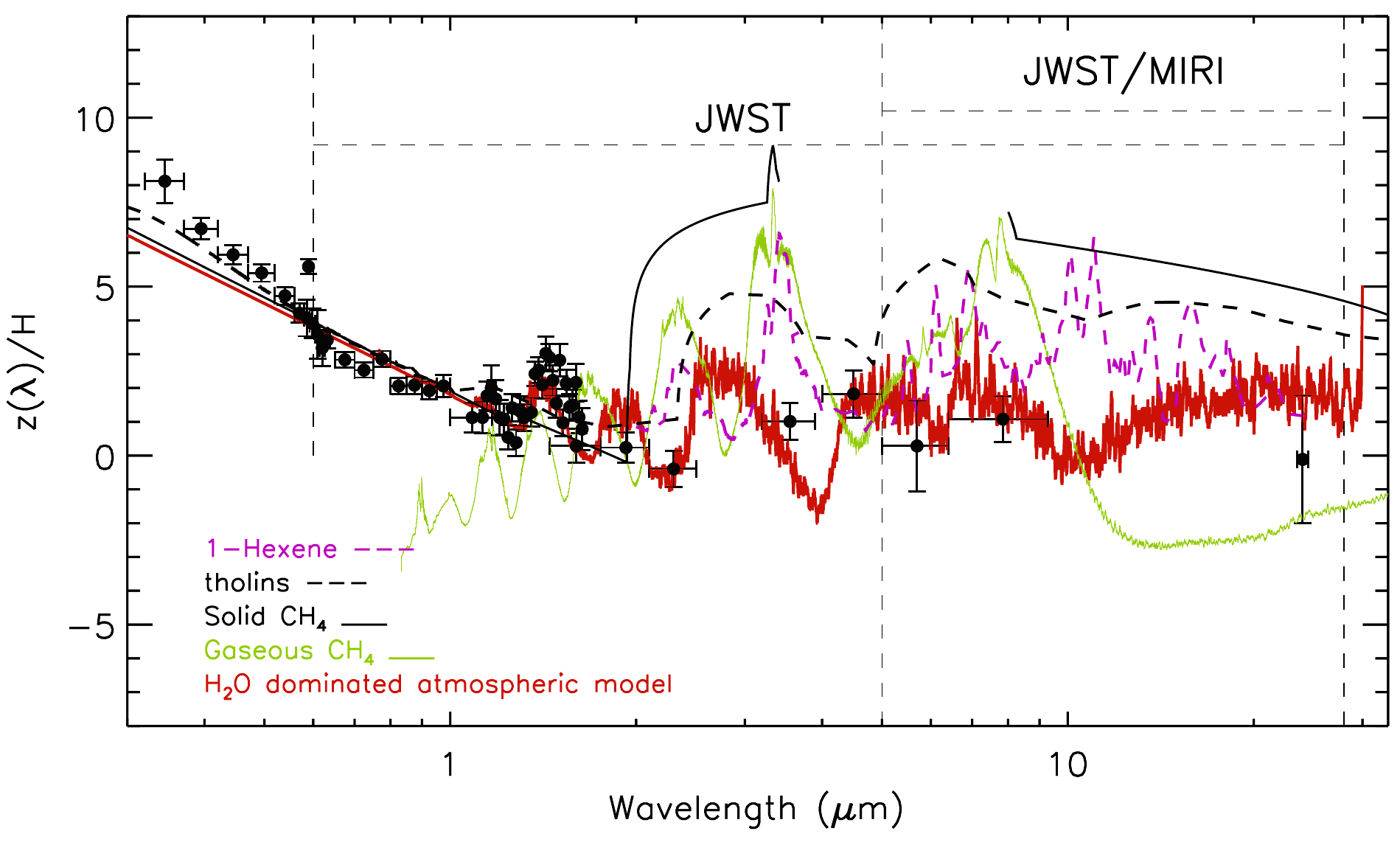}
\caption{Transmission spectrum of HD\,189733b (\citealt{pont2013}; \citealt{mccullough2014}) compared to hydrocarbon condensates and gaseous CH$_{4}$ (\citealt{yurchenko2014}; \citealt{amundsen2014}) and C$_{2}$H$_{4}$ (\citealt{rothman2009}; \citealt{sharp2007}) }
\label{fig:189_ch}
\end{centering}
\end{figure*}

Fe--O: The major vibrational mode features of iron oxides are centred at $\sim$14 with both broad- and narrow-band features from $\sim$9\,--\,40\,$\mu$m. For small grain sizes like that shown in Fig. \ref{fig:189_h2o} the transmission spectrum of Fe--O compounds exhibit a flattened slope from 0.3\,--\,2\,$\mu$m which could obscure the expected molecular water bands at 1.1 and 1.4\,$\mu$m. 
While potential iron-oxide narrow-band features, like those of Fe$_2$O$_3$, may be visible above the expected H$_{2}$O dominated transmission spectra (see Fig. \ref{fig:bond_plots}), it is unlikely that broadband features will be observed at the mbar pressure level where spectra dominated by gaseous vibro-rotational bands are several scale heights above any potential absorption features. 

Ti--O: Figure \ref{fig:189_h2o} shows the computed transmission spectrum of CaTiO$_{3}$, selected to represent potential Ti--O condensate clouds formed in hot Jupiter atmospheres. Ti--O molecules have vibrational mode features from 16\,--\,66\,$\mu$m which is partially covered by JWST/MIRI. While the cooler TiO$_{2}$ condensate shows little to no distinct features (see Fig. \ref{fig:bond_plots}), the hotter condensate CaTiO$_{3}$ has visible narrow absorption feature centred at $\sim$15\,$\mu$m and $\sim$21\,$\mu$m, which span several scale heights above the 1000$\times$ Fortney model.  

Sulphides: Sulphur-bearing compounds are expected to condense in hot Jupiter atmospheres at $\sim$1100\,K with highly scattering properties shortward of $\sim$10\,$\mu$m. Similar to iron oxide condensates, the transmission spectra shows a leveling off between 0.5 and 1\,$\mu$m which can potentially be used to help identify such clouds in UV and optical transmission spectra. In addition, the hotter condensates such as Al--O- and Fe--O-bearing compounds are not expected to coincide with sulphide condensate clouds, which helps identification. However, any additional features distinct to sulphur-bearing compounds do not emerge above the H$_2$O dominated model in hot Jupiter transmission spectra within the wavelength limits of JWST. 

Chlorides: Alkali chlorides begin to condense at around 800\,K, depleting the expected Na and K atomic species in the planets upper atmosphere. Figures \ref{fig:bond_plots} and \ref{fig:189_h2o} show that these condensates are highly scattering well into the infrared following a Rayleigh slope to $\sim$10\,$\mu$m. Vibrational modes are unlikely to be observed in hot Jupiter transmission spectra due to the obscuring molecular features, similar to sulphur-bearing condensate species. 
However, there is potential for chloride clouds to be inferred in cooler exoplanetary atmospheres where strong Rayleigh scattering is observed in the optical and there is no evidence for atomic gaseous Na or K in the optical spectra, suggesting that the species may have condensed out of the atmosphere forming clouds of liquid or solid particles.

Spectral measurements well into the infrared with extended wavelength resolution have the potential to differentiate between condensate species for a wide range of exoplanetary atmospheres as most of the cloud species have their lowest opacities at these mid-infrared wavelength regions. The 24\,$\mu$m point, of HD\,189733b further aids in the interpretation of the planets atmosphere placing strong constraints on the particle size given the optical vs. infrared absorption level, also noted by \citet{Lee2014}. While these vibrational modes of major bond species can be used to help identify the condensate cloud in the exoplanetary atmosphere, it is hard to ascertain the specific condensate responsible for the absorption feature present. Figure \ref{fig:189_h2o} shows cloud spectra extending above that of gaseous species, with a majority of potentially identifiable species having one or more absorption features in the wavelength range covered by JWST.

Identifying the different vibrational modes of potential cloud condensates and constraining\,/\,comparing the species in different exoplanet atmospheres could provide valuable insight into condensation chemistry over large temperature ranges. While the current broadband photometry is not of sufficient spectral resolution to distinguish between the models considered here, the situation should rapidly change with JWST. 

\begin{figure*}
\begin{centering}
\includegraphics[width=15cm]{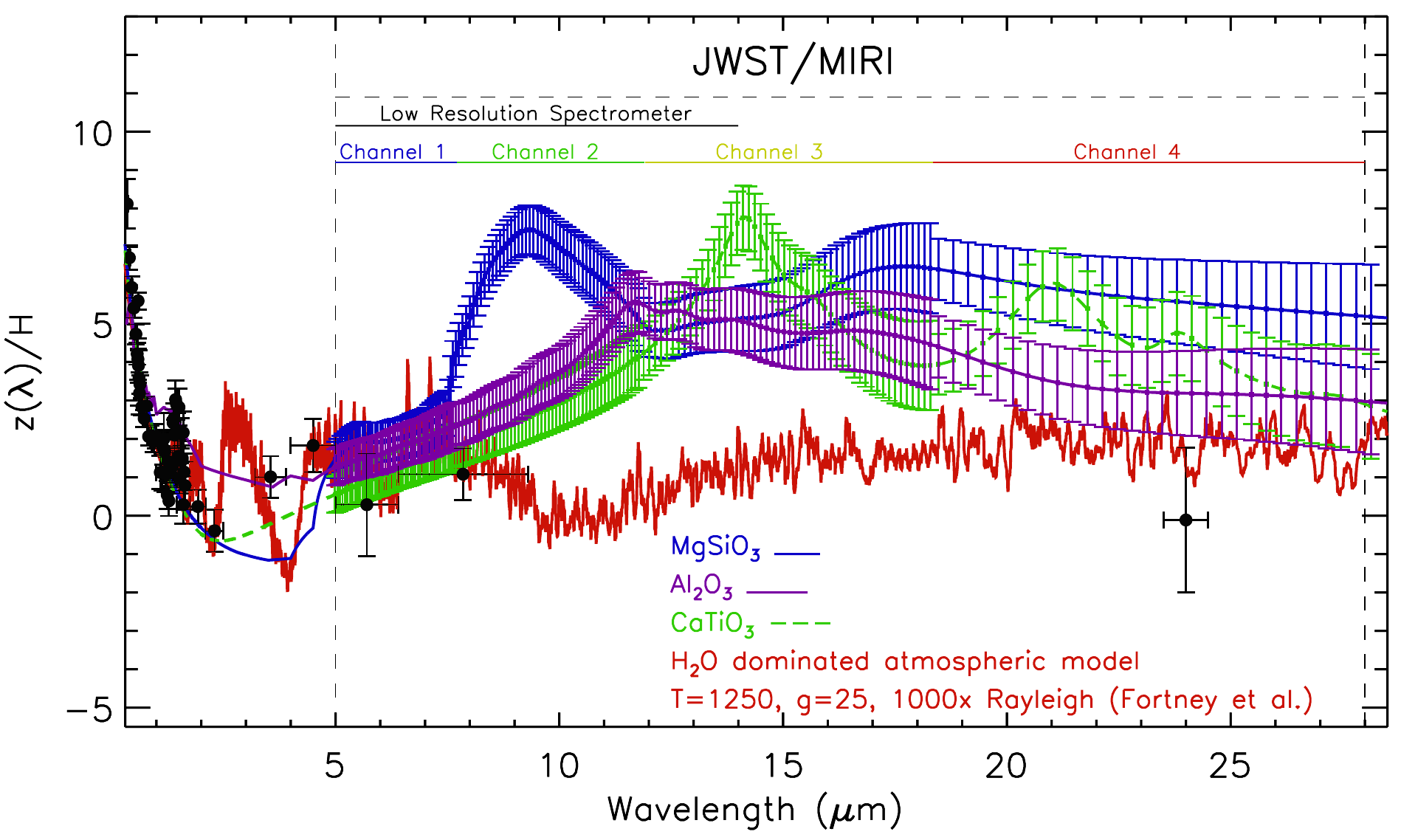}
\caption{Transmission spectrum of HD\,189733b (\citealt{pont2013}; \citealt{mccullough2014}) over-plotted on cloud model spectra set to R$\sim$50 for JWST/MIRI Medium resolution spectrograph channels 1--3 and R$\sim$30 for channel 4 with the wavelength coverage highlighted above the different regions of the spectra they cover. We also highlight the region of the spectrum covered by the low resolution spectrometer (5\,--\,14\,$\mu$m).}
\label{fig:189_res}
\end{centering}
\end{figure*}

%
\subsubsection{Photochemical Vs. Condensation}

Although unlikely to be present in the atmosphere of HD\,189733b, we have also included clouds composed of hydrocarbon species, such as hexene and Titan tholins, which may be generated photochemically. It can be seen that, although a poor fit to the HD\,189733b data, the major absorption features for condensate hydrocarbons extend above the 1000$\times$ Fortney model at the $\sim$3\,$\mu$m wavelength of the C--H vibrational mode and thus may be potentially observed for other planets.

In cooler planetary atmospheres, photochemistry is expected to play a key role in the overall cloud composition in the generation of gaseous hydrocarbons. When the planetary C/O ratio is greater than 1, the abundance of carbon-bearing compounds increases significantly, with disequilibrium processes enhancing their abundance over that of other species (\citealt{moses2013}).
Figure \ref{fig:189_ch} shows the transmission spectrum of HD\,189733b with the considered hydrocarbon species as well as the expected transmission spectrum for gaseous CH$_{4}$ at solar abundance. The opacities for gaseous CH$_{4}$ are calculated from the new ExoMol line list (\citealt{yurchenko2014}), with the line width parameters as in \citet{amundsen2014}. It can be seen that the gaseous CH$_{4}$ transmission spectral features overlap considerably with that of the hydrocarbon condensate cloud spectra, specifically at the 3\,$\mu$m range where the vibrational mode of the C--H bond is responsible for the absorption feature. There are, however, still some notable differences which may be used to differentiate between the two states with additional absorption features shown in both Hexene ($\sim$10\,$\mu$m) and Tholin ($\sim$6\,$\mu$m) condensate spectra which emerge several scale heights above both the H$_{2}$O dominated exoplanet spectra and that of the photochemically generated gaseous CH$_{4}$ (see Fig. \ref{fig:189_ch}).

%
\subsection{James Webb Space Telescope}

We have considered the different condensates that are expected to form condensate clouds in hot Jupiter exoplanet atmospheres with specific consideration to the wavelengths covered by James Webb Space Telescope (JWST). JWST, set to launch in October 2018, is a NASA mission, with significant contributions from both ESA and CSA. JWST is a 6.5\,m near- to mid-infrared telescope that will orbit at the Sun-Earth L2 point giving it an uninterrupted view of the sky. JWST is equipped with low, medium, and high resolution spectrographs from the instruments NIRSPEC (0.6\,--\,5\,$\mu$m) and MIRI (5\,--\,28\,$\mu$m). 

MIRI detectors are similar to Spitzer IRAC 5.8 and 8.0\,$\mu$m with an expected noise floor less than 100\,ppm. It is the only JWST instrument that will observe wavelengths greater than 5\,$\mu$m. The medium resolution spectrometer (MRS) is composed of four channels from 5\,--\,28\,$\mu$m with a resolution $\sim$3000--1000. To obtain a full transmission spectrum with MRS from 5--28\,$\mu$m four separate transit observations are required to observe in all four channels equaling $\sim$24 hours of observations for a majority of known exoplanet targets. 

Here we use HD\,189733b as an example hot Jupiter to simulate the transmission spectrum of our cloud condensate models and interpret the results with respect to the estimated precision\footnote{These estimates can be improved upon with an officially released specific exposure time calculator and increased understanding of the instrument systematics.} and resolution of the instruments used. 
A systematic noise floor value of 50\,ppm was adopted for the simulations following the framework outlined in \citet{fortney2013} with additional photon noise estimated using the MRS throughput (\citealt{Glasse2010}).
As discussed in Section \ref{sec:grainsize}, the presence of a strong optical slope in the transmission spectrum of HD\,189733b makes it a primary candidate for condensate cloud detection in the wavelength regime covered by JWST.
Figure \ref{fig:189_res} shows three of our cloud condensate models plotted to a resolution of R$\sim$50 across channels 1\,--\,3 and R$\sim$30 in channel 4.  
The spectra are binned significantly in each of the MIRI channels to increase the photon count at each wavelength therefore reducing the uncertainty of each wavelength bin. 

MIRI is ideal to detect the condensate vibrational mode features given in Table \ref{bond_freq} where most condensates have distinguishable features which can rise above the expected levels of the gaseous molecular features, with sulphide and chloride condensates a notable exception.
Using the example condensates shown for the atmosphere of HD\,189733b, transmission spectral observations using the two central channels (2\,\&\,3) of MIRI/MRS could be vital in distinguishing clouds formed from different condensate species.
Each of the Si--O-, Al--O- and Ti--O- bearing condensate compounds considered have absorption features extending several scale heights above that of the H$_2$O dominated molecular model which can be detected and resolved with MIRI. 
Chanel 2 of the MRS effectively covers the vibrational mode peak of Si--O- bearing compounds with absorption features several scale heights above that of other condensate species shown. Ti--O vibrational mode features are effectively covered by channel 3 of the detector.
Al--O- bearing condensate distinction would require observations over multiple channels of MRS, such that species such as Si--O and Ti--O can be ruled out and the infrared absorption features also matched to the levels observed in the optical. 

It is also possible to use the MIRI low resolution spectrometer (LRS) which covers 5\,--\,14\,$\mu$m with a R$\sim$100 (\citealt{fortney2013}). LRS encompasses both the major Si--O vibrational absorption mode at $\sim$9\,$\mu$m and absorption features generated by photochemical species such as tholins at $\sim$6\,$\mu$m. LRS is also advantageous as it requires only a single transit event to cover the entire wavelength range.
However, to identify vibrational modes from condensates at wavelengths longer than $\sim$14\,$\mu$m, MRS will be required.

%
%
\section{Summary} \label{sec:conclusion}
Clouds are now an increasingly important feature in many hot Jupiter atmospheres.
We have investigated the broad spectral properties of clouds, using Mie theory and analytic transmission spectral relations.   

We investigate the impact of grain size and distributions on condensate absorption spectra finding that the transmission spectrum can be well represented by the largest particle size in the distribution. Additionally, when a strong optical slope is observed in the optical, condensate vibrational mode features become prominent in the infrared associated with clouds composed of small sub-micron sized particles. 

Distinguishing cloud composition could in principle help make the distinction between cloud species generated photochemically or through condensation chemistry. We have highlighted spectral features in the infrared generated by the vibrational mode of a condensates major species bond as potential identifiers of cloud compositions in exoplanet atmospheres with the potential for both altitude and wavelength differentiation between species group.  
While it is difficult with current observations ($<$1.7\,$\mu$m) to distinguish between different cloud species we find cloud absorption features, caused by the vibrational mode of the major bond species of the condensate, could be present in the infrared, which could help discern different cloud types and constrain particles sizes and altitudes.  

The vibrational modes of the various condensates considered in this paper for hot Jupiter atmospheres span a large wavelength regime well into the infrared, with a majority between 9\,--\,28\,$\mu$m where current instruments cannot make transmission measurements. 

Of particular interest for this case study is MIRI, JWST's Mid InfraRed Instrument. 
MIRI has both imaging and spectroscopic capabilities from 5\,--\,28.3\,$\mu$m and will have 50 times the sensitivity and seven times the angular resolution of Spitzer, making it a vital instrument for the detection and characterisation of exoplanetary atmospheres as its long wavelength spectral capabilities will be highly sensitive to the cloud properties of transiting exoplanets.

%
%
\section{Acknowledgments} 
HRW acknowledges support from the UK Science \& Technology Facilities Council (STFC). The research leading to these results has received funding from the European Research Council under the European Union's Seventh Framework Programme (FP7/2007-2013) / ERC grant agreement n$^{\circ}$ 336792.
DKS acknowledges support from STFC consolidated grant ST/J0016/1.
The authors thank the anonymous referee for their useful feedback and comments.
We would also like to thank D.S. Amundsen, B. Drummond, J. Barstow, and K. Heng for useful discussions.

%
%
\footnotesize{
\bibliographystyle{mn2e}
\bibliography{theory}
}

\label{lastpage}
\end{document}